# Filamentary Extension of the Mem-Con theory of Memristance and its Application to Titanium Dioxide Sol-Gel Memristors

Ella Gale, Ben de Lacy Costello and Andrew Adamatzky
Unconventional Computing Group
University of the West of England
Bristol, UK

*Abstract*- Titanium dioxide sol-gel memristors have two different modes of operation, believed to be dependent on whether there is bulk memristance, i.e. memristance throughout the whole volume or filamentary memristance, i.e. memristance caused by the connection of conducting filaments. The mem-con theory of memristance is based on the drift of oxygen vacancies rather than that of conducting electrons and has been previously used to describe bulk memristance in several devices. Here, the mem-con theory is extended to model memristance caused by small filaments of low resistance titanium dioxide and it compares favorably to experimental results for filamentary memristance in sol-gel devices.

## I. INTRODUCTION

The memristor joins the resistor, capacitor and inductor as a fundamental circuit element. For symmetry reasons, it was predicted to be a passive two-terminal device that relates magnetic flux, $\varphi$, to charge, $q$ [1]. Because of the interaction of charge and flux, the memristor is the simplest non-linear circuit element and it possesses a memory. The memristor is considered to be a good candidate for neuromorphic computing due to the fact that biological synapses can be described as memristive systems (a memristor with two rather than one state variable) [2]. It also has possible applications in computer memory [3] and low-power consumption (green) computing as the device only draws power when changing state.

The archetypal and first reported memristor is the Strukov memristor, which is a titanium dioxide memristor made by atomic deposition [3]. This device had a non-linear I-V curve and memristor-like properties. A break-through in ease of manufacturability of the memristor was the announcement of a solution processed sol-gel memristor also based on titanium dioxide [4]. Both these devices worked by the movement of oxygen vacancies within the material, leading to a chemical change from the high resistance $TiO_2$ to low resistance $TiO_{(2-x)}$. The exact mechanism and structure of $TiO_{(2-x)}$ is unknown, but Magnéli phases have been implicated [5] (along with others [6]). Currently, both a bulk change of the $TiO_2$ material [3] and a small area filament change of material [7] have been suggested.

Since Strukov's report [3], there have been three theoretical approaches to describing real world memristors. The first was Strukov's phenomenological model [3] which is based on the movement of a boundary between the two types of titanium dioxide and which lacked magnetic flux (the magnetic flux would be expected as it is part of the definition of the device). In [3], it was stated that it is the non-linear I-V curve that defines the memristor, not its constitutive relationship (as given in [1]). Nonetheless, the lack of magnetic flux term has led to controversy where it has been questioned whether a true Chua memristor had been made [8]. In response, it was suggested that all resistance changing memories are memristors [9], a stance that would implicitly include all Resistive Random Access Memories, ReRAM. Several simulations have been performed which show memristor properties using either Strukov's model (for example Ref. [10]) or Chua's equations (for example, Ref. [11]).

The second theoretical model rewrote Strukov's equations as Bernoulli equations [12], thus suffering the same missing flux issue.

Only the third theoretical model, the mem-con model [13] resolved the problem of the magnetic flux. In the mem-con model [13], the resistance of the system is described as the sum of the memory and conservation functions, hence the name. Reference [13] identified a limitation with the previous phenomenological models where the charge movement which causes boundary movement was misidentified as electronic charge rather than ionic (oxygen vacancy) charge. The memory function of the Mem-Con model builds up a novel theory of memristance from magnetostatics by relating the vacancy magnetic flux to the total vacancy charge on the device. This method gives rise to a Chua memristance that satisfies Chua's constitutive relation as given in [1]. The Conservation function describes the time-dependent action of the rest of the device (the un-doped titanium dioxide).

In this paper we will extend the mem-con model to filamentary memristors and compare the theoretical results with experimental data.

## II. TYPES OF TITANIUM DIOXIDE SOL-GEL MEMRISTOR BEHAVIOUR

These devices work as memristors without the need for a forming step. We have noticed that without attempting to control the selection of device characteristics, titanium dioxide sol-gel memristors exhibit two types of behavior [14].

The first is a slow-changing I-V loop where both the high resistance state, $R_{off}$, and the low resistance state, $R_{on}$, are the same order of magnitude; this is the 'curved' switching type memristors. This type of switching behavior is encapsulated by current memristor theoretical models and is believed to be a result of a bulk change of the entire titanium dioxide volume as suggested in [3].

The second type of memristive behavior is a switch to a $R_{on}$ state orders of magnitude below the $R_{off}$ state; this is the 'triangular' switching behavior. This switching behavior is a more sudden switch with far more separated states than the slow changing loop. It demonstrates similarity to ReRAM unipolar switching, which is suggested to be due to the connection of filaments [7]. As the connection is to a much higher conduction state and happens at different voltages in successive I-V curve measurements, we suspect that this type of switching caused by the connection and disruption of filaments. Even if our devices are not filamentary, there are filamentary memristors and ReRAM devices which are not currently theoretically described. Memristors are often made out of similar components to ReRAM and the difference between the two devices is not a settled question.

The connection and disruption of filaments in a device are not directly included in current memristor models and, as there are many such devices, there is a clear need to set up a theoretical model for them.

## II. SETTING UP THE MODEL

The sol-gel memristor devices have titanium dioxide layer of thickness $D$ which is 40nm. Filaments in ReRAM or memristors are branching tree structures within conical frustrum envelopes [15,16]. Such filaments have been observed in TiO$_2$ ReRAM devices of the same thickness [17], our devices are likely more amorphous but as a first approximation the dimensions were taken from [17]. The system is modeled as a conical frustrum that fills up with vacancies creating a lower resistance phase of titanium dioxide, see Fig. 1. We are working with the model that all the memristance change in the device is filamentary based. Once the conical frustrum is full, the filament connects which causes a low (several orders of magnitude lower) resistance state.

The system can be represented as an equivalent circuit diagram, see Fig. 2. Note that this circuit is a model used in the theory, the experimental results reported in this paper are from real sol-gel memristors. The un-switchable bulk, $R_u$, colored white in Fig. 1, could be calculated from the geometry, but as it does not change, it is best considered as a type of contact resistance. The high resistance part of the conical frustrum is $R_{off}$, and this is in series with the low resistance part of the conical frustrum, $M$, which the Chua memristance as described for the conducting electrons. This is in parallel with the filament resistance, $R_{fil}$, for two reasons, firstly, even when filament is connected, some small amount of current still flows through the bulk and secondly, there still needs to be bulk drift of vacancies to allow the filament to disconnect. In terms of the mem-con model, the $M$ term is the memory function, and the $R_u$, $R_{off}$ are part of the conservation function. $R_{fil}$ represents an extension of the model to allow a switchable conducting filament.

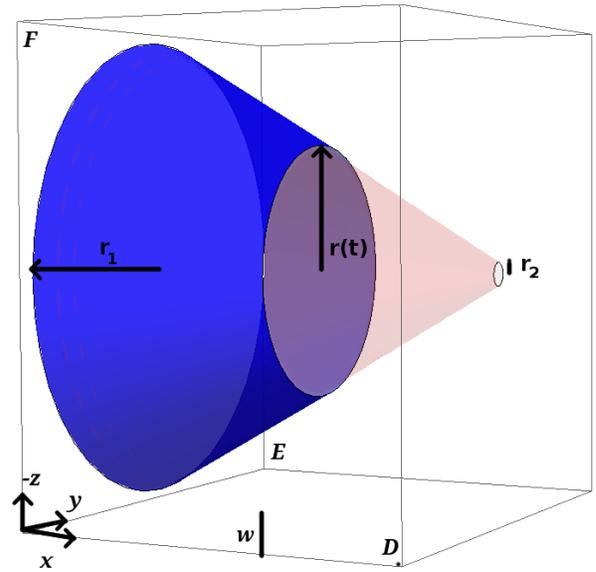

Fig. 1. Schematic for the filamentary memristor. The conical frustum envelope is shown within the volume of TiO$_2$ of the device (as given by $D \cdot 2E \cdot 2F$), note that this is not to scale, the real filaments are not on the same order as the TiO$_2$ volume. The radii $r_1$ and $r_2$ are to scale. The blue section represents the volume of TiO$_2$ given by $M$, the pink volume is $R_{off}$. As the filamentary envelope fills with lower resistance TiO$_2$, $r(t)$ changes. $R_u$ is represented by the white volume.

### THE RESULTING TOTAL RESISTANCE

The total resistance for this system, $R_{tot}$, is given by

$$R_{tot} = \frac{1}{\frac{1}{(R_u + M + R_{off})} + 2\,H(w-D)\frac{1}{R_{fil}}}, \qquad (1)$$

where $H$ is the Heaviside function. The Heaviside function acts as a switch to allow the filament to connect only when the boundary between doped and un-doped titanium dioxide, $w$, is at the end of the device, as the Heaviside function of a negative argument is zero. The Heaviside function is multiplied by two so that it is equal to 1 when $H(0)$. This term can be tuned to allow the filament to fuse before that point if desired. Equation 1 comes directly from the circuit schematic in Fig. 2. We will now briefly go through each term in turn.

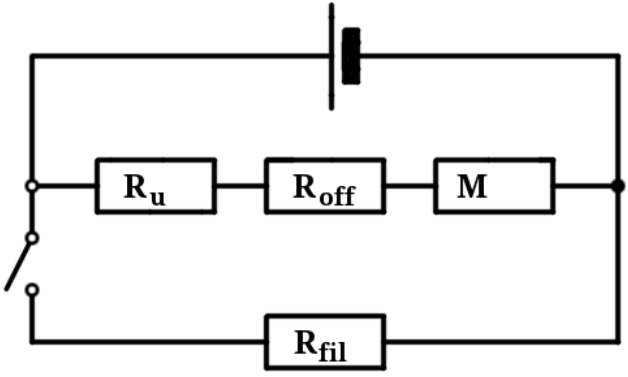

Fig. 2. Illustrative circuit diagram for the presented theoretical model. *M* is the memory function, the part of the TiO₂ that is in the $R_{on}$ resistance state. $R_{off}$ is the part of the TiO₂ that is in the off resistance state, and together with the un-switchable resistance, $R_u$, it makes up the conservation function. $R_{fil}$ is the resistance of the filament when it is connected and the connection is modeled by the switch.

## THE CHUA MEMRISTANCE

The hardest term to calculate in (1) is the memory function, *M*. To get this we need to calculate the Chua memristance, $M_c$, by calculating the magnetic flux, $\varphi$, that arises from the oxygen vacancy charge flow, *q*. To get this we calculate the magnetic field due to the ion vacancy current and integrate that over a surface to get a measure of the magnetic flux through that surface.

*A. The Conical Frustum Envelope Volume*

As the mem-con is a three-dimensional theory and therefore the geometry of the filaments needs to be carefully considered. This term covers the volume of the conical frustum envelope containing flowing vacancies, colored blue in Fig. 1.

We take the direction of ion vacancy flow as the +*x* direction. We consider a conical frustum envelope with vacancies drifting across it, where *w(t)* is the boundary between the self-doped material TiO₂ and the stoichiometric form. At this point, the cross-section of the frustum is a circle of radius *r(t)*. The vacancies grow from a large cross-sectional area of radius $r_1$ to a smaller one of radius $r_2$ which are 24nm [17] and 1.2nm respectively. From simple geometry, we get the volume of doped material, $Vol_{on}$, as

$$Vol_{on} = \tfrac{1}{3} \pi w \left(r_1^2 + 2 r_1 r(t) + r(t)^2\right) \quad (2)$$

and thus the volume of the stoichiometric material, $Vol_{off}$, is given by

$$Vol_{off} = \tfrac{1}{3} \pi w \left(r(t)^2 + 2 r(t) r_2 + r_2^2\right). \quad (3)$$

From these expressions and the volume for the entire conical frustum, we get an expression for *r(t)* in terms of *w(t)* and the dimensions of the frustum:

$$r(t) = \tfrac{1}{2D}\left(\frac{-2 w r_1 - 2 D r_2 + 2 w r_2 +}{\sqrt{n^2 + 4D(D r_1^2 - w r_1^2 + D r_1 r_2 + w r_2^2)}}\right)$$

where

$$n = 2 w r_1 - 2 D r_2 + 2 w r_2. \quad (4)$$

*B. Calculating the Magnetic Field Due to Oxygen Ions*

The volume current of vacancies, **J**, is the product of charge density per unit volume and the drift velocity, which can be written as

$$\mathbf{J} = (q(t)\, \mu_v\, \mathbf{L}) / Vol_{on}, \quad (5)$$

where $\mu_v$ the ionic mobility of the vacancies, L is is the electric field which causes their drift (in this device it is only non-zero in its *x*-component) and *q(t)* is the charge on the memristor due to the vacancy current: this is the *q* used in the definition of the Chua memristance. We will calculate the field at the edge of a box of dimensions *D* by *2E* by *2F*, where *E* and *F* are the *y* and *z* components of $r_1$. The time-dependent Cartesian components of the conical frustum's cross-sectional area, *e(t)* and *f(t)*, are calculated from $r(t)^2 = e(t)^2 + f(t)^2$. Equation 1 is then recast as a function of *e(t)* and *f(t)* and then substituted into (5), so we have **J** in terms of *e(t)* and *f(t)*, and this form is not included for reasons of brevity.

To calculate the magnetic field **B**, due to the oxygen ion flow, we use

$$\mathbf{B} = U \int G \, d\tau = \frac{\mu_0}{4\pi} \int \frac{\mathbf{J} \times \mathbf{r}}{r_x^3 + r_y^3 + r_z^3} \, d\tau, \quad (6)$$

where $\mathbf{r} \equiv \{r_x\, \mathbf{i},\ r_y\, \mathbf{j}, r_z\, \mathbf{k}\}$ and **i**, **j** and **k** are unit vectors in the Cartesian directions, $\mu_0$ is the permittivity of a vacuum and $d\tau$ is the volume infinitesimal within our cuboid $D \cdot 2E \cdot 2F$.

Solving the volume integral *G* can be done by parts ie. $\int G\, d\tau = uv \int v du\, d\tau$, where

$$u = \frac{1}{(r_x^2 + r_y^2 + r_z^2)^{3/2}}, \quad (7)$$

$$dv = \mathbf{J} \times \mathbf{r}, \quad (8)$$

and then *du* and *v* are found by differentiation and integration respectively, as is standard in this technique. The solution is

$$\mathbf{B} = U L \mu_v q \left\{ 0,\ -xz P_y(t)\, \mathbf{j},\ xy P_z(t)\, \mathbf{k} \right\}, \quad (9)$$

where $P_y$ and $P_z$ are based entirely on the geometry of the doped part of the system, and thus are dependent on *w*, which is dependent on *q* and *t*. The resulting field has a divergence of zero, as expected, and loops anti-clockwise around the *x* axis.

## C. The Value of the Magnetic Flux $\varphi$

To calculate $\varphi$ we must calculate the surface integral of magnetic flux over a relevant surface. As the magnetic **B** field has a zero *x* component, we cannot pick a surface parallel to *y-z* plane. To calculate the flux we use

$$\varphi = \int_{-E}^{+E} \int_{-F}^{+F} \mathbf{B} \cdot \mathbf{dA}\, dx\, dy, \qquad (10)$$

where **dA** is the surface normal for the infinitesimal area *dxdy*, as this points in the +*z* direction, the *z* component of the **B** field remains. The maximum value of $\varphi$ for a filament of these dimensions is $6.52 \times 10^{-30}$Wb, a tiny value, this is approximately a factor of ten less than the value for Strukov's memristor, and $1.65 \times 10^{-8}$ the value for bulk resistance of a sol-gel memristor device of the same dimensions [13]. This small value is due to the small volume of a filament and the fact that fewer oxygen ions are needed to form a filament than switch the bulk volume.

From Chua's constitutive equation, $\varphi = M_C(q)\, q$, [1] where $M_C$ is the Chua memristance, we get

$$M_C(q(t)) = U\, X\, \mu_v\, P_z(q(t)), \qquad (11)$$

where $X$ is the experimental constants, $2E \cdot 2F \cdot L$, and $2E \cdot 2F$ is the surface the flux was calculated over.

As $M_C$ is the memristance as experienced by the vacancy charge carriers we need the effective change in resistance for this volume of the device as experienced by the electrons, the Memory function, *M*. This is a different value as the electrons have a different mobility, $\mu_e$, in the material. However as both the electron mobility and the ion mobility are experimentally determined values, the memory function is given by

$$M = C\, M_c, \qquad (12)$$

where *C* is an experimentally determined constant. We have results from experimental data for this. However, we expect that this value arises from the physics of the device and it is currently under theoretical investigation.

## THE RESISTANCE DUE TO THE SWITCHABLE BUT UNSWITCHED VOLUME, $R_{OFF}$

The $R_{off}$ term in (1) is the resistance of the part of the conical frustum that is not doped, this is colored red in Fig. 1. Finding the resistance of a conical frustum has been widely used as an educational problem due to its apparent simplicity. However there are subtleties to the derivation which have been well explained [18]. In [18] the resistance of a conical frustum with spherical caps is reported, and as this result is only 3% off the numerically solved value (the standard textbook answer is 9% off), we shall use it here. Thus, for our system

$$R_{off} = \frac{\rho_{off}}{2\pi\, r(t) r_2} \frac{(r(t) - r_2)^2}{\sqrt{(D-w)^2 + (r(t) - r_2)^2} + (D-w)} \qquad (13)$$

Where $\rho_{off}$ is the resistivity of the high resistance part of the conical frustum. When *w* is very close to *D*, this term decreases in an unphysical manner and thus care must be taken.

## THE FILAMENT RESISTANCE

The $R_{fil}$ term is the resistance of the filament when connected and this is much lower than the on resistance and is, most likely, the cause of the much higher conduction state in 'triangular' switching memristors. The filament is actually a fractally-branched three-dimensional tree within the conical frustum envelope. Thus, although we can model the movement of oxygen vacancies across the conical frustrum, once it connects we shall model the actual filament. Reference [16] describes the fractal growth of a switching filament in titanium dioxide and gives the filamentary resistance as $R_{fil} = r_1^{-D_f + 1}$, where $D_f$ is 2.54 and we take $r_2$ as being small.

## COMPARISON TO EXPERIMENT

Figure 3 shows the results for the bulk memrisance. The Mem-Con theory reproduces the Lissajous curve as it should [1] (bottom curve). The top curve shows an example of bulk memristance in the titanium dioxide sol-gel devices. The experiment and the theory both have open loops, although the real world devices are more pinched at low voltages. This is advantageous as it is possible to read the devices at low voltage without significantly changing the device state.

Figure 4 shows the results of extending the model to include a filamentary mechanism, the top curve is the experimental data and the bottom is the theory. The theoretical model is a much closer fit to the experiment, which shows that we have the model of filament connection and disconnection works well. Just modeling filaments gives a good approximation to the experimental behavior, however, as figure 4 (top) shows, there is some non-linear bulk behavior happening when the device is in the off state.

These results allow us to make some comments about whether a device should be classified as a memristor or ReRAM. The Chua memristor is a device whose resistance changes in relation to the charge passed through the device [1]. The mem-con theory offers the clarification that this charge is the oxygen vacancy charge. Bulk bipolar memristance switching can be described by the memory function in (12) and it shows a continual change in resistance as the device charges (the pinched range in the middle shows that the ratio of this change is not constant over the entire voltage range). The conservation function for this type of device exhibits memristance as a result of the effect of the oxygen vacancies and thus, this type of device fit Chua's definition for a memristor.

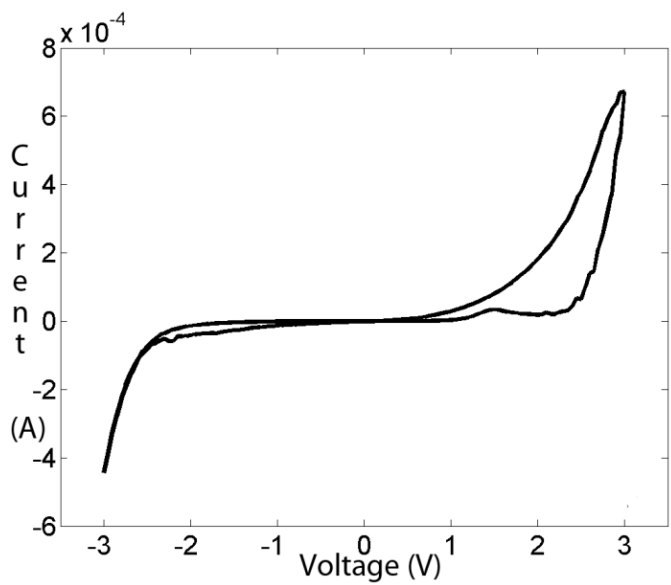
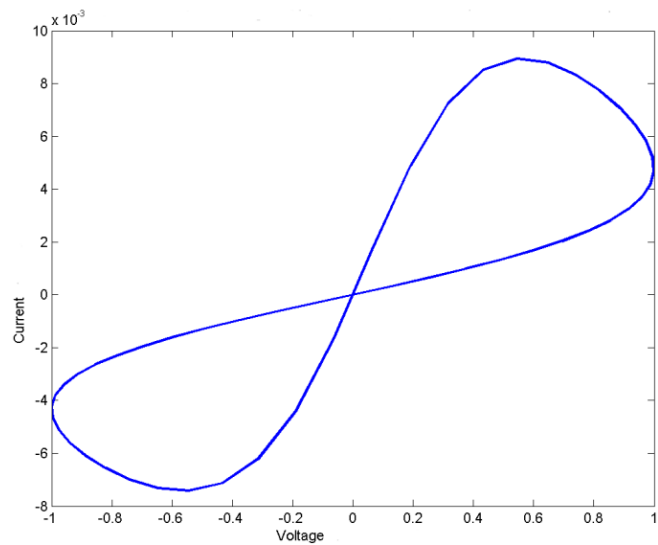

Fig. 3. Experiment (top) and theoretical model (bottom) for the bulk memristance switching. Note that the theoretical I-V curve is in reduced units. The theoretical curve is the similar in form to that expected from the phenomenological model and has the same qualitative shape as the measured I-V curve.

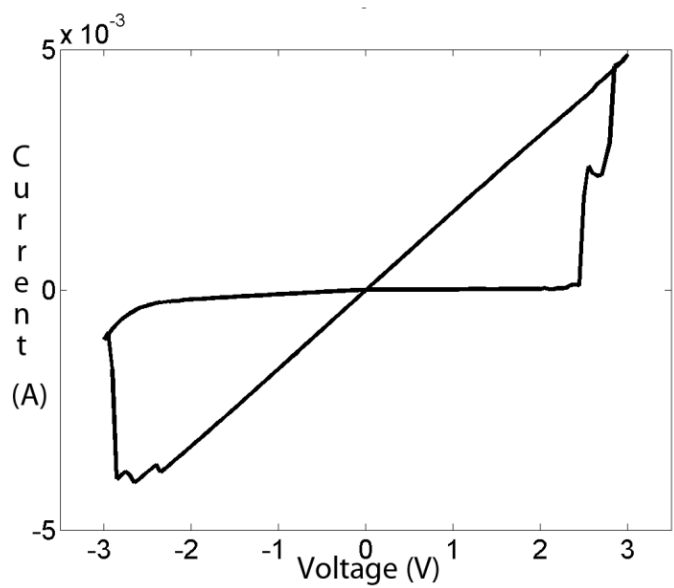
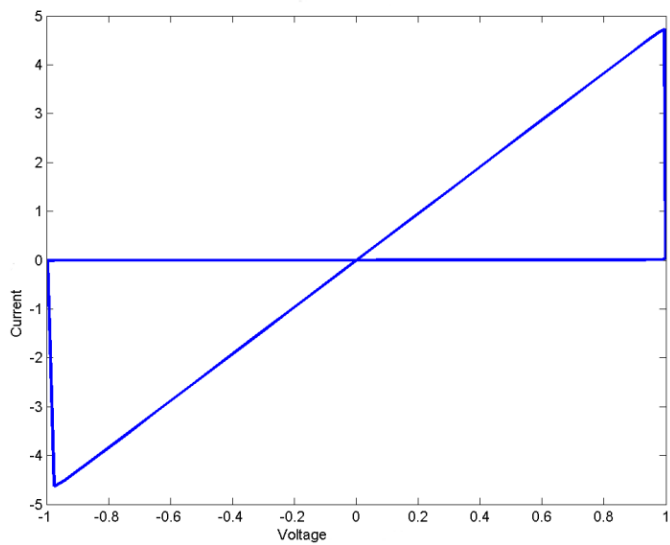

Fig. 4. Experiment (top) and theoretical model (bottom) for the filament memristance switching. Note that the theoretical I-V curve is in reduced units. There is a good match between the theory and experiment in terms of qualitative behavior.

To better describe the action of the triangular type of device we have added in a connectable and disruptable filament. When the filament is connected the device's current is ohmic (i.e.. linear with voltage) and its resistance does not change with time as a result of flowing vacancy charge. This represents a different mode of operation. Thus, this type of device is an extension of the Chua memristor. When in the high resistance state, the device exhibits Chua memristance, which is largely irrelevant when the filament is connected. However, as the overall I-V curve is a hysteretic loop, the device is still a resistance-switching memory. We postulate that filamentary memristors and, as they are very similar, ReRAM undergoing filamentary based unipolar switching, are best considered as extensions of Chua memristors rather than purely Chua memristors. They could be described as memristive systems where the second state variable is the connection state of the filament.

CONCLUSIONS AND FURTHER WORK

We have presented experimental data for two types of sol-gel devices and have extended the Mem-Con model of memristance to describe filamentary memristors. Previous work has described the memristive response when the vacancies are spread out throughout the bulk of the material [13]. The memristors described in this paper are only ones that operate by filaments, the oxygen vacancies are clustered together over a small area to create conducting filaments. This is an extreme case and real devices will also show some background bulk memristance, in fact, it may be that all memristors, like ReRAM, can be operative by either bulk switching or filamentary switching, dependent on the forming process the device is taken through and the voltage range it is driven over. This model can be combined with the original mem-con description by adding in the bulk memristance that happens in the part of the device current described by $R_u$.

The $R_{off}$ term decreases smoothly (but not linearly) as $w \rightarrow D$, but increases at the very end when $D - w$ is small, this leads to erroneous large current from this term. The solution to this is to change the boundary conditions from those presented here so that the filament connects at the point where the $R_{off}$ term is unphysical. This is a better physical model as we expect that the filament will connect sooner as a result of dielectric breakdown. Nonetheless, the overall model as presented here is a good first approximation to the system, as the current from $R_{fil}$ term of the equation drowns out any erroneously high current from the $R_{off}$ term at this point.

This work was undertaken to model our sol-gel devices and thus the next step is to work out the experimental constants to fit this model to experimental data. We intend to utilize such models to simulate, test and refine memristor logic circuit ideas alongside a program of experimental tests.

ACKNOWLEDGMENT

A. Author would like to acknowledge support from some people.